\documentclass[runningheads]{llncs}
\usepackage[T1]{fontenc}
\usepackage[utf8]{inputenc}
\usepackage{csquotes}
\usepackage{hyperref} 
\usepackage{amsmath}
\usepackage{amsfonts}
\usepackage{graphicx}
\usepackage{subfig}
\usepackage{listings}
\usepackage{siunitx}
\usepackage{xcolor}
\usepackage{cite}
\usepackage{multirow}

\lstdefinestyle{myStyle}{
    breaklines=true,
    frame=lines,
    numbers=none,
    captionpos=b,
    basicstyle=\scriptsize\ttfamily,
    keywordstyle=\bfseries\color{green!40!black},
    commentstyle=\itshape\color{purple!40!black},
    identifierstyle=\color{blue},
    backgroundcolor=\color{gray!5!white},
    escapeinside={<}{>}
}
\hypersetup{
    colorlinks=true,
    linkcolor=blue,
    filecolor=magenta,      
    urlcolor=blue,
    citecolor=blue,
    }

\def\mfldpy{\texttt{multi-freq-ldpy}}
\def\Mfldpy{\texttt{Multi-freq-ldpy}}    

\usepackage{geometry}
\begin{document}
\sloppy

\title{Multi-Freq-LDPy: Multiple Frequency Estimation Under Local Differential Privacy in Python\thanks{Authors are listed by order of contribution. Version of Record (ESORICS 2022): \url{https://doi.org/10.1007/978-3-031-17143-7_40}.}}

\titlerunning{Multi-Freq-LDPy: Multiple Frequency Estimation Under LDP in Python}

\author{Héber H. Arcolezi\inst{1} \and%
Jean-François Couchot\inst{2} \and%
Sébastien Gambs\inst{3} \and \\
Catuscia Palamidessi\inst{1} \and%
Majid Zolfaghari\inst{1,4} %
}

\authorrunning{H.H. Arcolezi et al.}

\institute{Inria and École Polytechnique (IPP), Palaiseau, France\\
\email{\{heber.hwang-arcolezi, catuscia.palamidessi, majid.zolfaghari\}@inria.fr} \and
Femto-ST Institute, Univ. Bourg. Franche-Comt\'e, UBFC, CNRS, Belfort, France\\ 
\email{jean-francois.couchot@univ-fcomte.fr} \and
Université du Québec à Montréal, UQAM, Montreal, Canada
\email{gambs.sebastien@uqam.ca} \and
Sharif University of Technology, Tehran, Iran
}

\maketitle              

\begin{abstract}
This paper introduces the \mfldpy\ Python package for multiple frequency estimation under Local Differential Privacy (LDP) guarantees. 
LDP is a gold standard for achieving local privacy with several real-world implementations by big tech companies such as Google, Apple, and Microsoft. 
The primary application of LDP is frequency (or histogram) estimation, in which the aggregator estimates the number of times each value has been reported. 
The presented package provides an easy-to-use and fast implementation of state-of-the-art solutions and LDP protocols for frequency estimation of: single attribute (\emph{i.e.}, the building blocks), multiple attributes (\emph{i.e.}, multidimensional data), multiple collections (\emph{i.e.}, longitudinal data), and both multiple attributes/collections. 
\Mfldpy\ is built on the well-established \textit{Numpy} package -- a \textit{de facto} standard for scientific computing in Python -- and the \textit{Numba} package for fast execution. 
These features are described and illustrated in this paper with four worked examples. 
This package is open-source and publicly available under an MIT license via GitHub (\url{https://github.com/hharcolezi/multi-freq-ldpy}) and can be installed via PyPi (\url{https://pypi.org/project/multi-freq-ldpy/}).

\end{abstract}

\section{Introduction} \label{sec:introduction}

Differential privacy (DP)~\cite{Dwork2006} is a formal privacy that allows to quantify the privacy-utility trade-off originally designed for the centralized setting. 
In contrast, the local DP (LDP)~\cite{first_ldp,Duchi2013} variant satisfies DP at the user-side, which is formalized as:

\begin{definition}[$\epsilon$-Local Differential Privacy]\label{def:ldp} A randomized algorithm ${\mathcal{M}}$ satisfies $\epsilon$-local-differential-privacy ($\epsilon$-LDP), where $\epsilon>0$, if for any pair of input values $v_1, v_2 \in Domain(\mathcal{M})$ and any possible output $y$ of ${\mathcal{M}}$:

\begin{equation*} \label{eq:ldp}
    \frac{\Pr[{\mathcal{M}}(v_1) = y]}{\Pr[{\mathcal{M}}(v_2) = y]}\leq e^\epsilon  \textrm{.}
\end{equation*}
\end{definition}

The privacy budget $\epsilon$ controls the privacy-utility trade-off for which lower values of $\epsilon$ result in tighter privacy protection. 
One fundamental task in LDP is frequency (or histogram) estimation in which the data collector (\emph{a.k.a.} the aggregator) decodes all the sanitized data of the users and can then estimate the number of times each value has been reported. 
The single frequency estimation task has received considerable attention in the literature (\emph{e.g.},~\cite{Bassily2015,tianhao2017,kairouz2016discrete,wang2016mutual,Min2018}) as it is a building block for more complex tasks dealing with temporal and/or multidimensional aspects.

More recently, in~\cite{Arcolezi2021_rs_fd} we have investigated the frequency estimation task of multiple attributes and proposed a solution named Random Sampling Plus Fake Data (RS+FD) that outperforms the state-of-the-art solution (divide users into groups to report a single attribute) commonly adopted in the literature ~\cite{xiao2,wang2019}. 
In addition, our work in~\cite{Arcolezi2021_allomfree} optimized state-of-the-art LDP protocols~\cite{kairouz2016discrete,rappor,tianhao2017} for longitudinal studies (\emph{i.e.}, multiple frequency estimation over time), which are based on the \textit{memoization} framework from~\cite{rappor}. 

In this paper, we introduce \mfldpy\footnote[1]{\url{https://pypi.org/project/multi-freq-ldpy/}}, which is the first open-source Python package providing an easy-to-use and fast implementation of state-of-the-art solutions and LDP protocols for the task of private multiple frequency estimation. 
By \enquote{multiple}, we mean either multidimensional data (\emph{i.e.}, multiple attributes)~\cite{xiao2,wang2019,Arcolezi2021_rs_fd}, longitudinal data (\emph{i.e.}, multiple collections throughout time)~\cite{rappor,microsoft,Arcolezi2021_allomfree}, or both multiple attributes/collections~\cite{Arcolezi2021_allomfree}. 
The package can be installed with PyPI using the pip command. 

\begin{lstlisting}[language=bash]
$ pip install multi-freq-ldpy
\end{lstlisting}

The \mfldpy\ package is mainly based on the standard \textit{numpy}~\cite{numpy} and \textit{numba}~\cite{numba} libraries, as the goal is to enable an easy-to-use and fast execution toolkit. 
The source code, documentation, several (Jupyter notebook) tutorials as well as an introductory video are available at the GitHub page (\url{https://github.com/hharcolezi/multi-freq-ldpy}). Released under the MIT open source license, \mfldpy\ is free to use and modify, and user contributions are encouraged to help enhance the library's functionality and capabilities.

\section{Presentation and Use Case Demo of multi-freq-ldpy} \label{sec:demonstration}

\Mfldpy\ is a function-based package that simulates the LDP data collection pipeline of users and the server. Thus, for each solution and/or protocol, there is always a \textit{client} and an \textit{aggregator} function. 
This section briefly presents the tasks that \mfldpy\ covers and presents four use-case of the library.

\subsection{Main Modules (Tasks Covered)} \label{sub:tasks}

The first task \mfldpy\ covers is \textbf{single-frequency estimation} under the \texttt{pure\_frequency\_oracles} module, which is a building block for the other tasks. The package currently features six\footnote[2]{A more complete Python package for \textit{single} frequency estimation can be found in (\url{https://pypi.org/project/pure-ldp/})~\cite{Cormode2021}.} state-of-the-art LDP protocols, namely: Generalized Randomized Response (GRR)~\cite{kairouz2016discrete}, Binary Local Hashing (BLH)~\cite{Bassily2015,tianhao2017}, Optimal Local Hashing (OLH)~\cite{tianhao2017}, Subset Selection (SS)~\cite{wang2016mutual,Min2018}, Symmetric Unary Encoding\footnote[3]{Originally known as basic one-time RAPPOR~\cite{rappor}.} (SUE)~\cite{tianhao2017}, and Optimal Unary Encoding (OUE)~\cite{tianhao2017}. 

Secondly, for \textbf{multidimensional frequency estimation} (\emph{i.e.}, multiple attributes), three solutions are implemented from~\cite{Arcolezi2021_rs_fd} with all aforementioned LDP protocols. These solutions, under the \texttt{mdim\_freq\_est} module, are: SPL) a naïve solution that splits the privacy budget $\epsilon$ over the total number of attributes; SMP) a state-of-the-art solution that randomly samples a single attribute and report it with $\epsilon$-LDP~\cite{wang2019,xiao2,Arcolezi2021_allomfree,tianhao2017}, and RS+FD) a state-of-the-art solution that randomly samples a single attribute to report with an amplified $(\epsilon' > \epsilon)$-LDP as it also generates one uniformly random fake data for each non-sampled attribute. 

Thirdly, for \textbf{single longitudinal frequency estimation}, \mfldpy\ features Microsoft's $d$BitFlipPM~\cite{microsoft} protocol and all the longitudinal LDP protocols developed in~\cite{Arcolezi2021_allomfree} based on the Google's RAPPOR~\cite{rappor} memoization solution (\emph{i.e.}, two rounds of sanitization). These protocols, following the \texttt{long\_freq\_est} module, are: Longitudinal GRR (L-GRR) that chains GRR in both rounds and four Longitudinal Unary Encoding (L-UE) protocols that chains SUE and/or OUE in both rounds of sanitization (\emph{i.e.}, L-SUE, L-SOUE, L-OUE, and L-OSUE). Indeed, L-SUE refers to the utility-oriented version of RAPPOR that chains SUE twice (\emph{a.k.a.} basic RAPPOR~\cite{rappor}). 

Finally, for \textbf{longitudinal multidimensional frequency estimation}, the package features both SPL and SMP multidimensional solutions with all the longitudinal protocols from~\cite{Arcolezi2021_allomfree} and Microsoft's $d$BitFlipPM~\cite{microsoft}, under the \texttt{long\_mdim\_freq\_est} module. 

\subsection{Worked Example: Single Frequency Estimation} \label{sub:single}

For example, the following use case demonstrates how easy it is to perform single frequency estimation with the \texttt{GRR}~\cite{kairouz2016discrete} protocol. 
In this example, there is a single attribute $A=\{a_1,...,a_k\}$ with domain size $k=|A|$, $n$ users, and the privacy guarantee $\epsilon$.
The complete code to execute this task is illustrated in Listing~\ref{fig:example_single} with the resulting estimated frequency for a given set of parameters and a randomly generated dataset. 
One can note that after the import functions, we essentially need two lines of codes to simulate the LDP data collection pipeline through applying the \texttt{GRR\_Client} and \texttt{GRR\_Aggregator} functions.

\begin{lstlisting}[language=Python, frame=tb, caption={Code snippet for performing single frequency estimation with the \texttt{GRR}~\cite{kairouz2016discrete} protocol.},float,label={fig:example_single}]
# Multi-Freq-LDPy functions for GRR protocol
from multi_freq_ldpy.pure_frequency_oracles.GRR import GRR_Client, GRR_Aggregator

# NumPy library
import numpy as np

# Parameters for simulation
eps = 1 # privacy guarantee
n = int(1e6) # number of users
k = 5 # attribute's domain size

# Simulation dataset following Uniform distribution
dataset = np.random.randint(k, size=n)

# Simulation of data collection
reports = [GRR_Client(user_data, k, eps) for user_data in dataset]

# Simulation of server-side aggregation
est_freq = GRR_Aggregator(reports, k, eps)
>>> array([0.199, 0.201, 0.199, 0.202, 0.199])
\end{lstlisting}

\subsection{Worked Example: Longitudinal Frequency Estimation} \label{sub:longitudinal}

In this second example, we demonstrate how to perform single longitudinal frequency estimation with the \texttt{L-SUE} protocol~\cite{Arcolezi2021_allomfree} (\emph{i.e.}, RAPPOR~\cite{rappor}) using \mfldpy. 
In this specific example, there is a single attribute $A=\{a_1,...,a_k\}$ with domain size $k=|A|$, $n$ users, and the privacy guarantees $\epsilon_{perm}$ (upper bound for infinity reports, \emph{a.k.a.} $\epsilon_{\infty}$ in~\cite{rappor}) and $\epsilon_1$ (lower bound for the first report\footnote[4]{Naturally, $0< \epsilon_1 \ll \epsilon_{perm}$ because higher values of $\epsilon_1$ are undesirable~\cite{rappor,Arcolezi2021_allomfree}.}). 
The complete code to execute this task is illustrated in Listing~\ref{fig:example_long} with the resulting estimated frequency for a given set of parameters and a randomly generated dataset. 

\begin{lstlisting}[language=Python, frame=tb, caption={Code snippet for performing single longitudinal frequency estimation with the \texttt{L-SUE}~\cite{Arcolezi2021_allomfree} (\emph{i.e.}, RAPPOR~\cite{rappor}) protocol.},float,label={fig:example_long}]
# Multi-Freq-LDPy functions for L-SUE (RAPPOR) protocol
from multi_freq_ldpy.long_freq_est.L_SUE import L_SUE_Client, L_SUE_Aggregator

# NumPy library
import numpy as np

# Parameters for simulation
eps_perm = 2 # longitudinal privacy 
eps_1 = 0.5 * eps_perm # first report privacy
n = int(1e6) # number of users
k = 5 # attribute's domain size

# Simulation dataset following Uniform distribution
dataset = np.random.randint(k, size=n)

# Simulation of data collection
reports = [L_SUE_Client(user_data, k, eps_perm, eps_1) for user_data in dataset]

# Simulation of server-side aggregation
est_freq = L_SUE_Aggregator(reports, eps_perm, eps_1)
>>> array([0.199, 0.201, 0.2, 0.198, 0.202])
\end{lstlisting}

\subsection{Worked Example: Multidimensional Frequency Estimation} \label{sub:multidimensional}

In another example, we demonstrate how to perform frequency estimation of multiple attributes with the RS+FD~\cite{Arcolezi2021_rs_fd} solution and the GRR protocol~\cite{kairouz2016discrete} using \mfldpy. 
In this setting, there are $n$ users, the privacy parameter $\epsilon$, and each user's profile is a tuple composed of $d$ attributes $\mathcal{A}=\{A_1, \ldots, A_d\}$ in which each attribute $A_j$ has a discrete domain of size $k_j=|A_j|$, for $j \in [1,d]$. 
The complete code to execute this task is illustrated in Listing~\ref{fig:example_multi} with the resulting estimated frequencies for a given set of parameters and a randomly generated dataset.

\begin{lstlisting}[language=Python, frame=tb, caption={Code snippet for performing multidimensional frequency estimation with the \texttt{RS+FD[GRR]}~\cite{Arcolezi2021_rs_fd} protocol.},float,label={fig:example_multi}]
# Multi-Freq-LDPy functions for RS+FD solution with GRR 
from multi_freq_ldpy.mdim_freq_est.RSpFD_solution import RSpFD_GRR_Client, RSpFD_GRR_Aggregator

# NumPy library
import numpy as np

# Parameters
eps = 1 # privacy guarantee
n = int(1e6) # number of users
k = 4 # attribute's domain size
d = 3 # number of attributes
lst_k = [k for _ in range(d)] # attributes' domain size

# Simulation dataset following Uniform distribution
dataset = np.random.randint(k, size=(n, d))

# Simulation of data collection
reports = [RSpFD_GRR_Client(user_tuple, lst_k, d, eps) for user_tuple in dataset]

# Simulation of server-side aggregation
est_freq = RSpFD_GRR_Aggregator(reports, lst_k, d, eps)
>>> array([0.255, 0.246, 0.248, 0.251], [0.252, 0.247, 0.249, 0.252], [0.252, 0.255, 0.244, 0.249])
\end{lstlisting}

\subsection{Worked Example: Longitudinal Multidimensional Frequency Estimation} \label{sub:long_mdim}

This last example demonstrates how to perform frequency estimation of multiple attributes (\emph{i.e.}, multidimensional data) throughout time (\emph{i.e.}, longitudinal data) with the SMP solution and the \texttt{L-GRR} protocol~\cite{Arcolezi2021_allomfree} using \mfldpy. 
In this setting, there are $n$ users, the privacy parameters $\epsilon_{perm}$ and $\epsilon_1$, and each user's profile is a tuple composed of $d$ attributes $\mathcal{A}=\{A_1, \ldots, A_d\}$ in which each attribute $A_j$ has a discrete domain of size $k_j=|A_j|$, for $j \in [1,d]$. 
The complete code to execute this task is illustrated in Listing~\ref{fig:example_long_multi} with the resulting estimated frequencies for a given set of parameters and a randomly generated dataset.

\begin{lstlisting}[language=Python, frame=tb, caption={Code snippet for performing longitudinal multidimensional frequency estimation with the SMP solution and \texttt{L-GRR}~\cite{Arcolezi2021_allomfree} protocol.},float,label={fig:example_long_multi}]
# Multi-Freq-LDPy functions for SMP solution and L-GRR protocol
from multi_freq_ldpy.long_mdim_freq_est.L_SMP_Solution import SMP_L_GRR_Client, SMP_L_GRR_Aggregator

# NumPy library
import numpy as np

# Parameters for simulation
eps_perm = 2 # longitudinal privacy 
eps_1 = 0.5 * eps_perm # first report privacy
n = int(1e6) # number of users
k = 4 # attribute's domain size
d = 3 # number of attributes
lst_k = [k for _ in range(d)] # attributes' domain size

# Simulation dataset following Uniform distribution
dataset = np.random.randint(k, size=(n, d))

# Simulation of data collection
reports = [SMP_L_GRR_Client(user_tuple, lst_k, d, eps_perm, eps_1) for user_tuple in dataset]

# Simulation of server-side aggregation
est_freq = SMP_L_GRR_Aggregator(reports, lst_k, d, eps_perm, eps_1)
>>> array([0.253, 0.255, 0.242, 0.250], [0.248, 0.253, 0.25, 0.249], [0.25, 0.245, 0.25, 0.255])
\end{lstlisting}

\section{Conclusion} \label{sec:conclusion}

In this paper, we have showcased the first open-source Python package named \mfldpy\ for private multiple frequency estimation under LDP guarantees. 
More specifically, we presented the modules of Version 0.2.4 of the library, but also its easy-to-use essence, requiring two lines of code to simulate the LDP data collection pipeline.
In addition to the standard single frequency estimation task, \mfldpy\ features separate and combined multidimensional and longitudinal data collections, \emph{i.e.}, the frequency estimation of multiple attributes, of a single attribute throughout time, and of multiple attributes throughout time. 
As an open source project, we welcome and encourage code contributions from the community to help grow and improve the library in all of its forms.
For future work, we aim to implement the LDP protocols of~\cite{Joseph2018,erlingsson2019amplification,Xue2022,Ohrimenko2022} for longitudinal data and those of~\cite{Varma2022} for multidimensional data. 

\section*{Acknowledgements}
This work was supported by the European Research Council (ERC) project HYPATIA under the European Union’s Horizon 2020 research and innovation programme. Grant agreement n. 835294. 
The work of Jean-François Couchot was supported by the EIPHI-BFC Graduate School (contract ``ANR-17-EURE-0002").
Sébastien Gambs is supported by the Canada Research Chair program as well as a Discovery Grant from NSERC.

\bibliographystyle{splncs04}
\bibliography{ms.bib}

\end{document}